\documentclass{optica-article}

\journal{opticajournal} 

\articletype{Research Article}

\usepackage{lineno}
\usepackage{threeparttable}
\usepackage{booktabs}
\newcommand{\DelEps}{$\Delta\epsilon$\ }
\newcommand{\showcase}{show-case}
\newcommand{\relUnc}{\sigma_{rel}}

\graphicspath{ {./figures/} }

\begin{document}

\title {Unambiguous determination of optical constants and thickness of ultrathin films by using optical anisotropic substrates}

\author{Sebastian Schaper (né Funke),\authormark{1,*} Matthias Duwe,\authormark{2} and Ursula Wurstbauer\authormark{1}}

\address{\authormark{1}Institute of Physics, University of Muenster, Wilhelm-Klemm-Str. 10, 48149 Muenster\\
\authormark{2}Park Sytems GmbH, Stresemannstr. 30, 37075 Goettingen}

\email{\authormark{*}sebastian.schaper@uni-muenster.com} 


\begin{abstract*} 
The unambiguous and universal determination of optical constants such as complex refractive indices (n, k) and thickness (d) in one measurement is typically confined for films with a thickness of more than 10~nm that hampers its application for ultra-thin films such as two-dimensional materials. We demonstrate that the commonly accepted limit of n, k, d coupling is overcome in ellipsometry by utilizing anisotropic substrates. In this way, ellipsometry allows to determine n, k and d simultaneously for thicknesses lower than 1~nm with high accuracy. Extensive simulations proof the potential of using anisotropic substrates for decoupling. The simultaneous determination of n, k and d is of great interest in several fields, two-dimensional materials, their heterostructures, ultra thin films for application in quantum technology, plasmonics and nano-photonics.

\end{abstract*}

\section{Introduction}
With the ongoing and still growing interest in two-dimensional (2D) materials and their heterostructures for application in opto-/electronics, and integrated photonics as well as other ultrathin films towards neuromorphics and quantum technology \cite{leiGrapheneRecentAdvances2022, pham2DHeterostructuresUbiquitous2022, kasparRiseIntelligentMatter2021, sebastianTwodimensionalMaterialsbasedProbabilistic2022, kwakSensorComputingUsing2023, mennelUltrafastMachineVision2020, migliatomaregaLargescaleIntegratedVector2023, wolffSuperconductingNanowireSinglephoton2020, kangExploitingPlasmons2D2022, pendurthiHeterogeneousIntegrationAtomically2022, stevesUnexpectedInfraredVisible2020, lundebergTuningQuantumNonlocal2017,bankwitzHighqualityFactorTa2O5oninsulator2023a}, we have the increasing need for unambiguous and simultaneous determination of optical properties and thickness even in the limit of mono- to few layers, particularly 2D-materials \cite{dasTransistorsBasedTwodimensional2021,fukamachiLargeareaSynthesisTransfer2023, radisavljevicSinglelayerMoS2Transistors2011, troueExtendedSpatialCoherence2023, wietekNonlinearNegativeEffective2024,dirnbergerMagnetoopticsVanWaals2023}. 
Imaging Ellipsometry (IE) is widely applied to characterize optical properties and identify layer number of 2D-materials \cite{grudininHexagonalBoronNitride2023,funkeImagingSpectroscopicEllipsometry2016,wurstbauerImagingEllipsometryGraphene2010, braeuninger-weimerFastNoncontactWaferScale2018}, their hetero- and hybridstructures \cite{siggerSpectroscopicImagingEllipsometry2022} as well as atomically thin 2D polar metals  \cite{rajabpourTunable2DGroupIII2021, nisiLightMatterInteraction2021} but also thin films used for integrated photonic and plamsonic circuitries \cite{bankwitzHighqualityFactorTa2O5oninsulator2023, nisiLightMatterInteraction2021}. 
Ellipsometry and imaging ellipsometry are non-contact optical methods that allow the unambiguous and universal measurement of the complex refractive indices $n+ik$ and thickness $d$ of thin-film layers. The thin film parameters ($n, k, d$) can be obtained simultaneously as long as the they decouple in the computational data analysis. However, this is not the case for ultra-thin films of about $d < 10~nm$, for which the so-called n, k, d-coupling prevents the unambiguous analysis of ellipsometric data unless a priori knowlegde of the optical properties or layer thickness is available\cite{fujiwaraSpectroscopicEllipsometryPrinciples2007}.
This coupling is often a huge limitation if e.g. new materials are investigated with layer dependent properties, a specific layer thickness as e.g. hexagonal Boron nitride (hBN) mono- or bilayers as tunnel barriers \cite{britnellElectronTunnelingUltrathin2012} are required or both - thickness and optical constants are changed by an external stimuli such as temperature\cite{parkTemperatureDependenceCritical2016,nguyenTemperatureDependenceDielectric2024} or electric field \cite{luInfluenceElectricField2006} e.g. in ultrathin films of piezoelectric\cite{martin-sanchezStraintuningOpticalProperties2017,plumhofStraininducedAnticrossingBright2011}, phase change or charge density wave phase materials\cite{leeOpticalPropertiesChargedensitywave2008}. 

While approaches exist to handle this  n, k, d coupling with specific solutions e.g. using absorbing layers \cite{hilfikerSurveyMethodsCharacterize2008}, transparent layers \cite{droletPolynomialInversionSingle1994}, 
additional measurements \cite{hilfikerSurveyMethodsCharacterize2008, ibrahimParameterCorrelationComputationalConsiderations1971},  complementary determination of the thickness \cite{guAnalyticalMethodDetermine2020} or advanced/complicated analytical approaches \cite{gilliotExtractionComplexRefractive2012}, all of those methods are very specific for certain material systems and are not universally applicable.

Here we propose a novel approach to break the n, k, d coupling by using anisotropic substrates in order to obtain optical properties and thickness for thin layers unambiguously even in the limit of mono- to few layers with a thickness below 1~nm. 

To demonstrate the conceptual power of our approach we selected various material combinations of thin film and substrate (both transparent, absorbing, metallic). We evaluate the n, k, d decoupling in ellipsometry for ultrathin films as a function of the substrate's anisotropy and thickness of the layer. We therefore simulate the ellipsometric data for realistic material combinations, where available, and evaluate the n,k, d coupling in the required numerical modelling approach in order to extract the relevant parameters from the simulated data. We start showcasing an extreme material combination: A thin layer of the transparent 2D-material hBN, on a transparent substrate to which we introduce an artificial anisotropy, resulting in an optical birefringence of the substrate. Further \showcase{}s will feature transparent, semi-transparent, and conductive layers on both transparent and semi-transparent substrates. These substrates will incorporate Rutile as a realistic birefringent substrate. 

\section{Method}

Ellipsometry measures the sample-induced change of the polarization state, which occurs upon the reflection of an electromagnetic (EM) wave from a sample surface. For isotropic non-depolarizing samples, this interaction can be described by the complex coefficients of reflection ($r_{pp}$, $r_{ss}$) for the linear p and s-polarization states, respectively. Ellipsometry measures the complex ratio $\rho$ thereof, whose amplitude and phase are traditionally denoted by the ellipsometric angles $\Delta$ and $\Psi$, respectively\cite{fujiwaraSpectroscopicEllipsometryPrinciples2007}:
\begin{gather}
\rho = \frac{r_{pp}}{r_{ss}} = tan\Psi * exp^{-i\Delta}.
\label{rho}
\end{gather}

For stratified media, $\Delta$ and $\Psi$ depend on the sample properties (e.g. material optical constants, layer thickness, surface roughness) as well as on the angle of incidence (AOI) and  the vacuum wavelength $\lambda$ of the probing EM wave. 

\begin{figure}[ht]
\caption{Orientation of coordinate system and scheme of sample structure used within this paper. The ambient is the surrounding medium and not explicitly indicated.}
\centering
\includegraphics[width=0.46\textwidth]{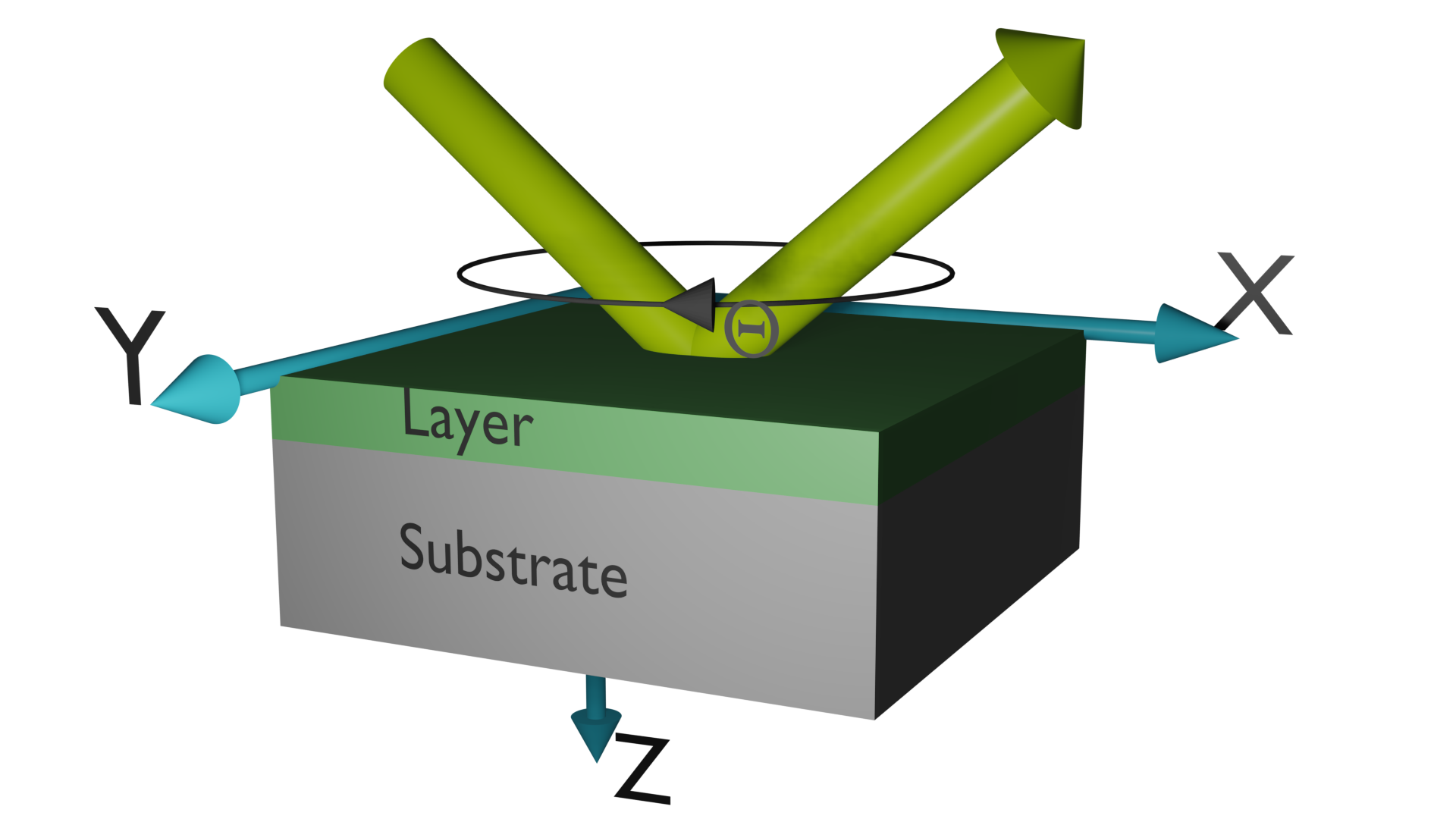}

\label{fig:sample}
\end{figure}

Ellipsometric data analysis solves the inverse problem that deduces the sample properties from the ellipsometric readout by means of computational modelling\cite{azzamEllipsometryPolarizedLight1987}.
A sample model is herein described by an ambient, through which the incident EM wave impinges the sample surface, by one (or more) layers of specific thickness $d$, and by a flat substrate that is considered to be semi-infinite, thus no further reflections from the backside are observed (Fig. \ref{fig:sample}). The optical parameters of the involved materials are denoted by the refractive index $n$ and extinction coefficient $k$ yielding the complex refractive index $N = n+ik$, that is related to the complex dielectric function via $N^2 = \epsilon$.\footnote{Within this paper we apply the physics sign convention for the complex dielectric function $\epsilon = \epsilon_1 + i\epsilon_2$ to describe the optical parameters, which is favourable for the implementation of the employed modelling algorithm. Please note that the sign of the phase term in eq.\ref{rho} inverts ($i\rightarrow-i)$ if the optics sign convention $\epsilon = \epsilon_1 - i\epsilon_2$ is employed.\cite{fujiwaraSpectroscopicEllipsometryPrinciples2007}} To solve the inverse problem, the free (unknown) model parameters are iteratively adjusted to find the best match of calculated and experimental data, which in general allows a simultaneous determination of optical parameters and layer thicknesses \cite{haugeDesignOperationETA1973, jellisonMeasurementOpticalFunctions1997}. 

However, the solution of the inverse problem becomes ambiguous if there is a high numerical correlation of the model parameters. This is the typical case for a single, ultra-thin  layer on a flat substrate, for which $\Psi$ carries almost no information about the film if $d \lesssim 10 \text{ nm}$, and $\Delta$ provides only the correlated information about the "optical thickness" $N_{layer}\cdot d$ \cite{fujiwaraSpectroscopicEllipsometryPrinciples2007}, which even persists if the measurement contains data points from multiple AOI and $\lambda$ (variable angle of incidence spectroscopic ellipsometry, VASE).

The previous considerations hold for samples that comprise isotropic materials only. However, the situation changes if the sample comprises anisotropic materials. For example, if we choose an uniaxial linear birefringent substrate with an off-cut optical axis (i.e. \textit{not} parallel to the z-axis in  Fig. \ref{fig:sample}), the substrate anisotropy induces p-to-s and s-to-p cross-polarization upon the interaction of the EM wave with the sample, resulting in a non-diagonal Jones matrix
$J =
\begin{bmatrix}
r_{pp} & r_{ps} \\
r_{sp} & r_{ss}
\end{bmatrix}$.

The non-zero parameters $r_{ps}$ and $r_{sp}$ denote the complex p-to-s and s-to-p cross-polarization coefficients, respectively. The Jones matrix now additionally depends on the azimuthal orientation of the sample $\Theta$. This dependency can be used to gain additional information about the thin-film by probing the sample at different orientations.

We investigated the impact of the substrate anisotropy on the solution of the inverse problem for ultra-thin films to decorrelate the information of $n,k$ and $d$ by means of numerical simulations. In the following, we introduce an uniaxial in-plane anisotropy to the dielectric tensor of the substrate material and implemented Berreman's 4x4 matrix description to calculate the sample's Jones matrix $J$ for an isotropic ambient, a single isotropic thin film, and an anisotropic substrate\cite{berremanOpticsStratifiedAnisotropic1972}. The sample structure with the film thickness, the materials' optical properties, as well as "measurement" parameters AOI, $\lambda$, and $\Theta$ are readily contained in the 4x4 matrix \cite{fujiwaraSpectroscopicEllipsometryPrinciples2007, schubertInfraredEllipsometrySemiconductor2004}.

The uniaxial substrate material is described by the ordinary (o) and extra-ordinary (eo) dielectric constants $\epsilon_o$ and $\epsilon_{eo}$. Applying the coordinate system shown in Fig. \ref{fig:sample}, the substrate's eo-axis shall coincide with the x-axis at $\Theta=0$. We thus define the diagonal elements ($\epsilon_{xx},\epsilon_{yy}, \epsilon_{zz}$) of the substrate dielectric tensor at $\Theta=0$ :
\begin{equation}    
    \begin{gathered}
        \epsilon_{yy} = \epsilon_{zz} = \epsilon_o 
    \end{gathered}
        \label{eq:exx}
\end{equation}

\begin{equation}
    \begin{gathered}
        \epsilon_{xx} = \epsilon_{eo} = \epsilon_o  - \Delta\epsilon
    \end{gathered}
    \label{eq:eyy}
\end{equation}
where \DelEps{} is the linear birefringence of the substrate. The azimuth $\Theta$ of the sample is defined clockwise (topview) around the z-axis, and the dielectric tensors at non-zero azimuths were calculated from the respective Euler rotation matrices \cite{fujiwaraSpectroscopicEllipsometryPrinciples2007}.

The simulations are carried out with the following Mueller-Jones matrices notation:
\begin{gather}
    M = A ( J \otimes J^*) A^{-1},
\end{gather}
as defined in Appendix 4 of \cite{fujiwaraSpectroscopicEllipsometryPrinciples2007}. For a generalized case we use Mueller-Jones matrices in our simulations, but it shall be noted that  $\Delta$ and $\Psi$ or generalized ellipsometry may be sufficient in the experimental realization, as the optical axis is rotating within the sample plane (xy) and no depolarization is assumed. This is particular important, because not all ellipsometer are capable to access the Mueller-Jones matrices of a sample.

In order to evaluate the impact of a substrate birefringence \DelEps{} on the degree of n, k, d-coupling of the thin film, we simulate the Mueller-Jones matrix $M_{ij,l}^{mod}$ ($i,j=1,...,4$) for a given set of substrate ($\epsilon_o$, \DelEps) and thin-film ($\epsilon_{layer}$\footnote{Or equivalently $n, k$.}, $d$) parameters at multiple sample azimuths $\Theta$ by using a Python implementation of Berreman's 4x4 method. We further emulate the experimental data by adding normal-distributed noise ($\sigma=1e^{-4}$) to the Mueller-Jones matrices, and then solve the inverse problem for $\epsilon_{layer}$, $d$ by numerical minimization of the mean square error (MSE):
\begin{gather}
    MSE = \sqrt{\sum\limits_{l=1}^{N}\sum\limits_{i=1}^{3}\sum\limits_{j=1}^{4}\left[\left(\frac{M_{ij,l}^{exp} - M_{ij,l}^{mod}}{\sigma_{M_{ij,l}^{exp}}}\right)^2\right] }
    \label{eq:MSE}
\end{gather}
where $N$ denotes the number of different sample azimuths. Please note that the defined MSE discards the 4th row of the Muller-Jones matrix, i.e. using only the top 3x4 submatrix. We chose 3x4 Mueller-Jones matrices on purpose, because it matches the measurement capabilities of the most common ellipsometer types\footnote{Most common ellipsometers are of the PCSA (or PSCA) type, which can only measure the 3x4 (or equivalently 4x3) Mueller sub-matrices. PSCA nomenclature describes the setup as Polarizer-Compensator-Sample-Analyzer.}.

The LMFit python package \cite{newvilleLMFITNonLinearLeastSquare2014} is used for non-linear regression analysis by using Levenberg-Marquardt algorithm to minimize the MSE and obtain the best fit results. We investigated the regression results by means of a uniqueness fit of the thickness parameter $d$. This means in the context of this work the thickness was iterated over a certain range, while $\epsilon_{layer}$ parameters were be changed by the regression analysis to minimize the MSE. The resulting MSE for each thickness is recorded and normalized to the lowest MSE obtained in the thickness range. In this way, the overall best fit values of $d$ and $\epsilon_{layer}$ is at the normalized $MSE=1$, and the uncertainty $\sigma_d$ of the  obtained best fit value of $d$  is given by the thickness range defined by $MSE\leq 1.1$ \cite{hilfikerDeterminingThicknessRefractive2017}. We use the relative uncertainty of the thickness 
\begin{gather}
    \relUnc = \frac{\sigma_d}{d}
    \label{eq:rel_sigma}
\end{gather}
as a figure of merit to quantify the impact of the substrate anisotropy on the uniqueness of the ellipsometric data analysis.

\begin{figure*}[ht]
\caption{Uniqueness fit of 1~nm (left), 3~nm (middle) and 10~nm (right) hBN layer on transparent substrate with real birefringence of Rutile (red line) and hypothetically artificially induced \DelEps{} (black lines), at an angle of incidence of 50\textdegree and at a wavelength of 500~nm. The bottom line shows zoomed-in regions to highlight the  change in the  MSE curves with the horizontal line  at 1.1 indicating the boundaries for the derived uncertainty during the uniquess fit  approach. }
\centering
\includegraphics[width=1\textwidth]{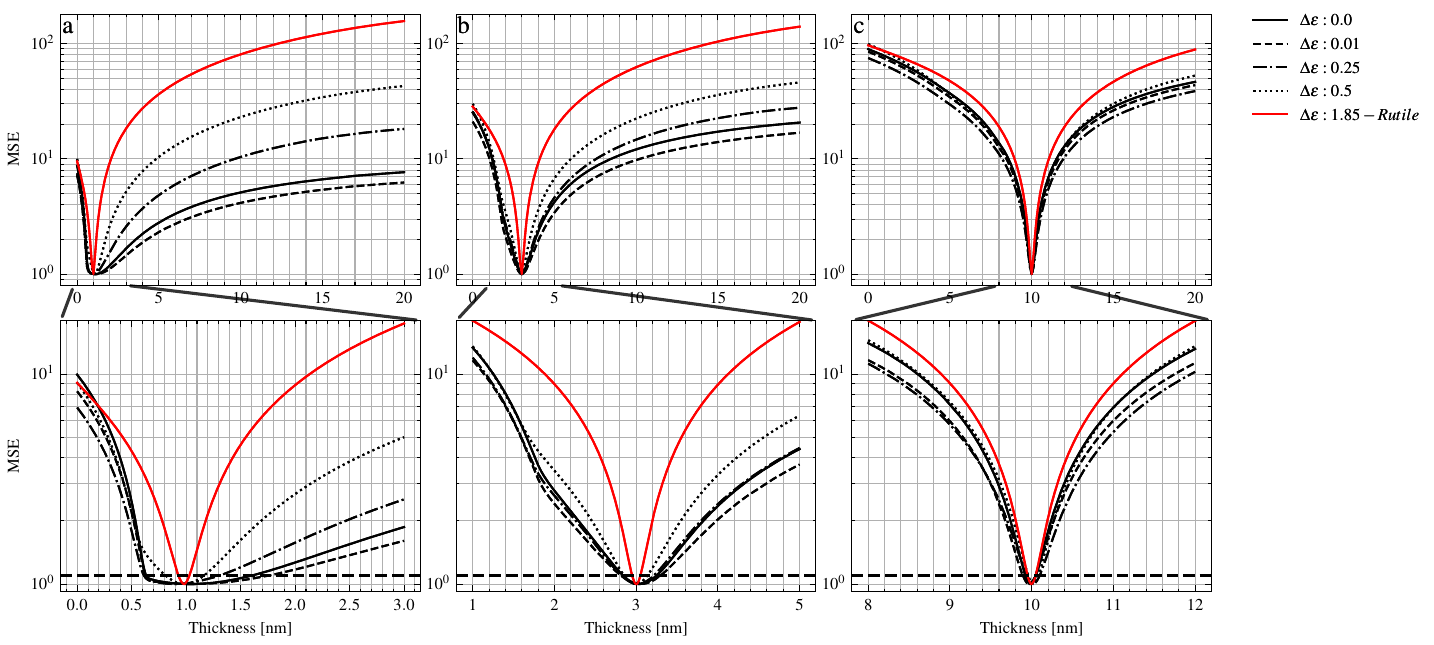}

\label{fig:uniqueness}
\end{figure*}~

\section{Simulations}
\subsection{hBN on Rutile}
\label{sec:hBNRutile}
In order to test a realistic scenario, we choose an ultrathin hBN-layer on a Rutile substrate. The simulation is done for a single wavelength of 500 nm at an AOI of 50\textdegree,  combining three sample azimuths $\Theta=[0,45,90^\circ]$ and using $\epsilon_{layer}= 5.0$, $\epsilon_o=7.4$ and $\Delta\epsilon=1.85$ as dielectric constants of hBN and Rutile, respectively   \cite{grudininHexagonalBoronNitride2023,devoreRefractiveIndicesRutile1951}. The uniqueness fits for the nominal hBN-layer thicknesses of 1, 3 and 10~nm are given by the solid red lines in Figure \ref{fig:uniqueness}(a-c). 
The MSE curves were obtained from eq.~\ref{eq:MSE} over a layer-thickness grid (0 to 20~nm, stepsize 0.05~nm) by non-linear regression of the dielectric constant $\epsilon_{layer}$. Each curve shows a distinct minimum at the nominal thickness values of 1, 3 and 10~nm as expected. The (...) [right column (b,d,f)] shows the same cases but with a higher resolution and limited to the range of $\pm$2~nm around the nominal layer thickness. It can be readily seen that for a Rutile substrate the shape of the uniqueness curves look almost identical for each case. Thus, the analysis yields similar uncertainty values $\sigma_d ^{MSE\leq1.1}$ of 0.2~nm (Figure~\ref{fig:uniqueness}d), 0.24~nm (Figure~\ref{fig:uniqueness}e), and 0.21~nm (Figure~\ref{fig:uniqueness}f), clearly undercutting the typical thickness limit of about 10~nm for the simultaneous extraction of layer thickness and dielectric constant. 

To test the sensitivity of this approach, we further assumed that the substrate's birefringence was virtually tunable in the range of \DelEps{} = 0 to 1.85 (black lines in Figure~\ref{fig:uniqueness}). The results for the change in the relative uncertainty $\sigma_{rel}$ according to eq.~\ref{eq:rel_sigma}
are summarized in  Table~\ref{tab:uncertaintieshBNRutile}.
For the 10~nm case (Figure~\ref{fig:uniqueness}c,f), the substrate's birefringence only has a minor impact on the MSE curves. The variation of \DelEps{} slightly changes the curvature of the MSE, but a clear minimum of the MSE is observed at the real thickness of 10~nm for all tested values of \DelEps. Even for the isotropic case (solid black line), the relative uncertainty is just $\sigma_d=0.7\%$ vs. the 0.4\% obtained for Rutile. 
The outcome is different for the ultra-thin layers.
In the 3~nm case (Fig.~\ref{fig:uniqueness}b,e), the MSE trace shows a broadened minimum for the low birefringence values (\DelEps<0.5). Accordingly, the relative uncertainty is almost 5 times larger in the isotropic case ($\sigma_{rel}=6.8\% $) than for the Ruitle substrate ($\sigma_{rel}=1.5\%$).
In the case of the 1~nm layer (Fig.~\ref{fig:uniqueness}c,f), the MSE traces and the uncertainty of the thickness are most sensitive to variation of \DelEps. The MSE traces feature a well-defined minima for the high birefringence values (\DelEps=0.5, 1.85), but not for the isotropic or near-isotropic cases. 
The relative thickness uncertainty drops from $48.4\%$ for the isotropic substrate to $4.9~\%$ (or in absolute numbers: 0.02~nm!) for the Rutile case. 
The cause for the increase of $\relUnc${} if \DelEps{} is  0.01 may be caused by this specific combination of dielectric constants of layer and substrate, but is not fully understood.
Our results demonstrate that the use of a Rutile substrate with its large linear birefringence may provide sufficient sensitivity for the simultaneous determination of the thickness and dielectric constant of ultra-thin hBN layers.

Within the field of 2D-materials the thickness of a layer is typically denoted by the inter-atomic distance in bulk material times the layer number and ellipsometry is applied to get the thickness or similar layer-number depending optical properties as shown in \cite{funkeImagingSpectroscopicEllipsometry2016, grudininHexagonalBoronNitride2023,okanoVoltageControlledDielectricFunction2020}.

\begin{table}[htbp]
\centering
\caption{\bf Relative uncertainties for thicknesses of 10~nm, 3~nm and 1~nm of hBN on Rutile with different \DelEps}
\begin{tabular}{c|ccc}
\hline
\DelEps{} &  & $\relUnc$ &  \\
\hline
  & 10~nm hBN & 3~nm hBN & 1~nm hBN \\
\hline
0 & $0.7\%$ & $6.8\%$ & $48.4\%$ \\
0.01 & $0.7\%$ & $8.2\%$ & $56.6\%$\\
0.25 & $0.7\%$ & $6.5\%$ & $34.5\%$\\
0.5 & $0.6\%$ & $4.7\%$ & $17.6\%$\\
1.85 & $0.4\%$ & $1.5\%$ & $4.9\%$\\
\hline
\end{tabular}
  \label{tab:uncertaintieshBNRutile}
\end{table}

\subsection{Simulation of various material combinations}

Different layer and substrate material combinations are investigated for their influence on the relative uncertainty. We define six \showcase{}s with different material combinations.
Table \ref{tab:usedEpsilons} summarizes the parameters used as input parameters for the simulations of the selected \showcase{}s including values of $\epsilon_{o}$ for the substrate and the $\epsilon_{layer}$ of the ultrathin layer. The dielectric values are taken from refractiveindex.info \cite{polyanskiyRefractiveindexinfoDatabaseOptical2024}, an open source database, for the following materials hBN\cite{grudininHexagonalBoronNitride2023}, Rutile\cite{devoreRefractiveIndicesRutile1951}, Silicon\cite{aspnesDielectricFunctionsOptical1983}, Graphene\cite{weberOpticalConstantsGraphene2010},  MoS\textsubscript{2}\cite{ermolaevBroadbandOpticalProperties2020} and Quartz\cite{ghoshDispersionequationCoefficientsRefractive1999}. The \showcase{}s are chosen to cover variations, of $\epsilon_{layer}$ being lower and larger than $\epsilon_o$ of the substrate. Both combinations are implemented for transparent and semitransparent thin layers  each. In addition, the materials for the layer are taken from 2D-materials as limit for ultrathin materials \cite{funkeImagingSpectroscopicEllipsometry2016, wurstbauerImagingEllipsometryGraphene2010} and $\epsilon$ is always considered isotropic for the layer material.
For the substrate, the dielectric values of the ordinary axis are taken from existing materials, but it should be noted, that for simulations, the birefringence \DelEps{} was again assumed to be virtually tunable (compare section \ref{sec:hBNRutile}). 

  We use the complex dielectric function of each thin film material and substrate as defined for the \showcase. The ambient material for all cases is air. Simulations are done for a wavelength of 500~nm at an AOI of 50\textdegree and 3 rotational angles of 0\textdegree, 45\textdegree and 90\textdegree. The ellipsometric data sets are generated by calculating 3x4 Mueller-Jones matrices for given input and add normal distributed noise of $1e^{-4}$. Afterwards the uncertainty of thickness and best parameters for thicknesses ($d_{Fit}$) and ($\epsilon_{Fit}$) are obtained by uniqueness fits, see Figure \ref{fig:uniqueness} and section \ref{sec:hBNRutile}. The creation and fit is repeated 5 times to average the uncertainty and cancel out variations caused by the noise. The fitting procedure starts at the expected values for $\epsilon_{layer}$, to minimize the simulation time. 
  For transparent layer materials the real part is changed during the routine, whereas for semi-transparent layer materials the real and imaginary part of $\epsilon$ are varied. To simplify matters, we concentrate on the thickness uncertainty, with the behaviour of $\epsilon_{layer}$ being presented in the supplemental material Figure S1 c). This behaviour closely resembles that of the uncertainty in thickness.

\begin{table}[ht]
\centering
\begin{threeparttable}
\caption{\bf Dielectric values for $\lambda=$500nm used for simulating the \showcase{}s I - VI. \expandafter\MakeUppercase  \showcase{}\ I is transparent layer on transparent substrate ($\epsilon_{layer} < \epsilon_{susbstrate}$), II for absorbing layer on transparent substrate, III transparent layer on transparent substrate  ($\epsilon_{layer} > \epsilon_{susbstrate}$), IV for transparent layer on absorbing substrate,V for absorbing layer on absorbing substrate and VI for metallic layer on absorbing substrate.}
\begin{tabular}{cccc}

\toprule
       \showcase{}& materials & $\epsilon_{layer}$ & $\epsilon_{o}$ \\
\midrule
I & hBN-Rutile\tnote{*} &        5.0+0.0j &   7.4+0.0j\\
II & MoS\textsubscript{2}-Rutile\tnote{*}  &      21.2+10.8j&   7.4+0.0j \\
III & hBN-Quartz\tnote{*$\dagger$}  &        5.0+0.0j &   2.4+0.0j\\
IV &    hBN-Si  &        5.0+0.0j&  18.5+0.6j \\
V &   MoS\textsubscript{2}-Si  &      21.2+10.8j &  18.5+0.6j\\
VI & Graphene-Si  &        5.8+6.4j &  18.5+0.6j\\
\bottomrule

\label{tab:usedEpsilons}
\end{tabular}
\begin{tablenotes}
\footnotesize
\item[*] uniaxial by nature
\item[$\dagger$] optical activity is neglected
\end{tablenotes}
\end{threeparttable}
\end{table}

\begin{figure*}[ht]
\caption{Simulation of relative layer-thickness uncertainty in $\relUnc$ of a thin layer on an uniaxial substrate for the cases hBN-Rutile (a), $MoS_2$-Rutile (b), hBN-Quartz (c), hBN-Si(d), $MoS_2$-Si (e) and Graphene-Si (f). Simulations are done at $\lambda = 500nm$; $AOI = 50^\circ$. The artificially induced birefringence of \DelEps{} and thickness of layers (hBN, $MoS_2$, Graphene) are varied to study the impact of the subtrate birefringence on the n, k, d decoupling using the relative thickness uncertainty $\relUnc$ as figure of merit.}
\centering
\includegraphics[width=1\textwidth]{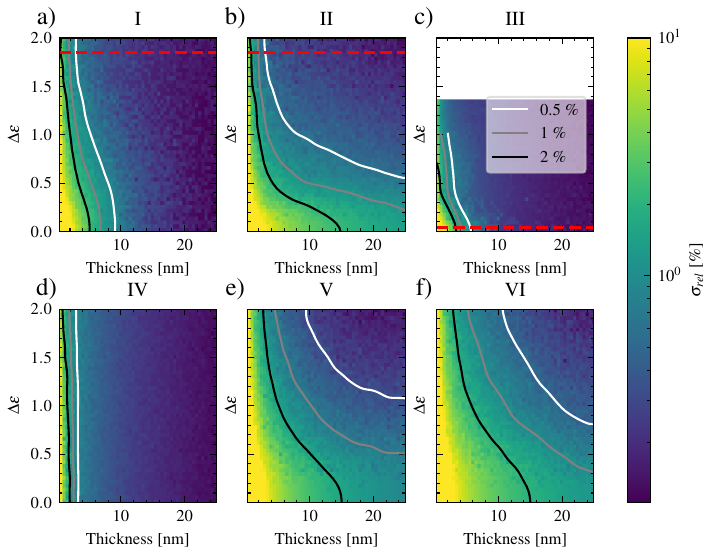}

\label{fig:simulations}
\end{figure*}

To visualize our findings, we varied \DelEps{} in 30 steps from 0.0 to 2.0 for each thickness, 30 steps in between of 0.1~nm to 25~nm, and show the $\relUnc$ of the thickness for each combination obtained by a uniqueness fit. Figure \ref{fig:simulations} a) shows such map of $\relUnc$ for the case of hBN on Rutile, our example from the previous section. For each combination of thickness and \DelEps{} the $\relUnc$ is simulated via a uniqueness fit as explained in detail in chapter \ref{sec:hBNRutile}. Therefore, $\epsilon_{layer}$ will be varied for fixed thicknesses within a certain thickness range. The best fit value and uncertainty of $\epsilon_{layer}$ and thickness are derived from the MSE minium.
The color scale in Figure \ref{fig:simulations} is logarithmic for a better visualization. Iso-curves are drawn for 0.5~\%, 1~\% and 2~\%. Table \ref{tab:isoCurves} summarizes the thicknesses $d_s$ for the pure isotropic substrate and a substrate with $\Delta\epsilon = 0.6$ for all \showcase{}s at 2~\% $\relUnc$. 2~\% is chosen, to allow a comparison of all \showcase{}s.

If $\epsilon_{layer}$ is lower than the substrate's $\epsilon_{o}$ (\showcase{}\ I), an increase in \DelEps{} decreases the uncertainty of thickness of the transparent layer (Figure \ref{fig:simulations} (a)). 
A significant overall reduction of the correlation between $\epsilon_{layer}$ and $d$ sets in for a substrate birefringence larger than 0.6.
The results for \showcase{} II, a semi-transparent layer (semiconducting MoS\textsubscript{2} on transparent substrate (insulating Rutile) are displayed in Figure \ref{fig:simulations} (b). The overall behaviour is similar to (\showcase{}\ I). However the overall decrease in uncertainty for $\Delta\epsilon = 0$ is lower. The higher $\relUnc$ might originate from the finite imaginary part of the dielectric function for MoS\textsubscript{2} that is optimized in the fit approach, too. An increase in \DelEps{} minimizes the uncertainty as it does for the \showcase{} I. 

Show-case III, displayed in Fig \ref{fig:simulations} (c) is for a transparent layer (insulating hBN) on a transparent substrate (insulating Quartz) with the $\epsilon_{layer} > \epsilon_o$ at the chosen wavelength of 500~nm. Variations of \DelEps{} $\geq$ 1.4 are not considered, because therefore the real part of the dielectric function of the extra-ordinary term of Quartz would be smaller than 1 while the imaginary part being zero. For this \showcase{}an overall reduction of uncertainty is observed. The uncertainty for the isotropic case is already lower than for the case where the $\epsilon_{layer}$  is smaller than the substrate. Red line indicates Quartz crystal. This \showcase{} in combination with \showcase{} I demonstrates that a change in absolute value of $\epsilon_o$, may change the absolute value $\relUnc$ but an increase in \DelEps{} would still decrease $\relUnc$.

\begin{table}[htbp]
\centering
\caption{\bf Thickness $d_s$ at $\relUnc$ of 2\% for isotropic and $\Delta\epsilon = 0.6$ substrate.}
\begin{tabular}{cccc}
\toprule
       \showcase{}& materials & $d_s$($\Delta\epsilon=0$) & $d_s$($\Delta\epsilon=0.6$)\\
       &  & [nm] & [nm]\\
\midrule
I & hBN-Rutile &   5.0 &        2.7 \\
II & MoS\textsubscript{2}-Rutile &   15.0 &      3.2 \\
III & hBN-Quartz &   3.2 &        0.9 \\
IV & hBN-Si &  1.7 &       1.6 \\
V & MoS\textsubscript{2}-Si &  15.1 &      8.2 \\
VI &Graphene-Si &  11.0&        9.5\\
\bottomrule
\end{tabular}
  \label{tab:isoCurves}
\end{table}

In the case of transparent material (hBN) on semi-transparent substrate (Si) (\showcase{}IV), see Figure \ref{fig:simulations} (d), the relative uncertainty is much lower compare to the other test cases for all values of \DelEps{}. Any increase in \DelEps{} only changes the relative uncertainty by a little. For example a 2~\% relative uncertainty can be seen in the isotropic case for 1.7~nm and in the case of \DelEps=0.6 for 1.6~nm (solid black line, Figure \ref{fig:simulations}).

For the cases of semi-transparent layer on semi-transparent substrate (\showcase{}s V, VI) the simulation results are shown in Figure \ref{fig:simulations} (e) and (f), respectively. Panel (e) displays the relative uncertainty for $\epsilon_{layer}$ being higher than $\epsilon_o$ of the substrate by simulating MoS\textsubscript{2} on top of silicon (Si) (\showcase{}\ V), whereas for Graphene on Silicon (\showcase{}\ VI, Figure \ref{fig:simulations} f), the complex dielectric function of the film is lower, than the substrate's dielectric function. The overall relative uncertainty increases for both cases compared to the other cases. For semi-transparent layers the real and imaginary part are fitted during the uniqueness fit, instead of only the real part for transparent layer materials. Nonetheless, a decrease in relative uncertainty can be seen for MoS\textsubscript{2}-Si and graphene-Si with increasing \DelEps{} of the substrate birefringence.

It shall be noted, that we choose a single wavelength and single AOI for the simulations for the sake of clarity. Many labs are equipped with single-wavelength ellipsometers and don't have access to spectroscopic measurements. In addition the single-wavelength and single AOI approach is a realistic use-case to use the ellipsometric enhanced contrast mode of IE, that allows to tune the contrast to make mono- to few-layer of 2D-materials visible \cite{braeuninger-weimerFastNoncontactWaferScale2018}. It is import to note, that further decoupling can be achieved using a multiple angle and spectroscopic approach. 
In general using a second AOI reduces the uncertainty as shown in supplemental material Figure S1, but the dependency on thickness and optical properties will be the same \cite{hilfikerSurveyMethodsCharacterize2008}. Rotational angles can be reduced down to 3 angles (see supplemental material Figure S2), without sacrificing the decoupling. The change in the absolute value of the substrates $\epsilon_{o}$ shows, that the achievable decoupling depends on the involved dielectric functions of the materials (Figure S5 of the supplemental material). 
The same results and decoupling effect observed when fitting for layer thickness can also be achieved by fitting the dielectric constants of the layers instead, as illustrated in Figure S3 of the supplementary materials. Additionally, the supplementary materials provide comprehensive insights into the error behaviour associated with the fitted $\epsilon_1$ value of the layer during a unique thickness fit, as depicted in Figure S4.
We find a decrease in the uncertainty of thickness for all simulated values for anisotropic substrates with \DelEps. We would like to note, that the \DelEps{} has to be in the substrate in order to achieve a decrease of uncertainty (Figure S6 of the supplemental material).

\section{Results and Outlook}
We have shown by thorough modelling, that an anisotropic substrate can be used to decouple optical properties from thickness information of thin layers in ellipsometry. Various \showcase{}s have been simulated and discussed. Overall, a reduction in the relative uncertainty of the extracted layer thickness is achieved by including an in-plane anisotropy of the substrate. However, the amount of reduction depends on the materials combinations. The results are promising for the field of ellipsometry. Additional measurement parameters, such as wavelength and angle of incidence, may reduce the uncertainty further. It may also open a new approach, by using metamaterial substrates e.g. \cite{liMicrowaveBirefringentMetamaterials2016, schmidtAnisotropicBruggemanEffective2013} to fine tune the optical properties of the substrate for the best decrease in uncertainty for thin materials and can also offers a pivotal role in the determination of the optical properties of 2D-materials. We want to emphasize the \showcase{}s I, II and III, where real anisotropic materials have been simulated. Rutile and quartz substrates show a promising decrease in uncertainty for transparent layers, e.g. hBN and semi-transparent materials, e.g. MoS\textsubscript{2}. Thus, for exfoliated flakes of 2D-materials with unknown $\epsilon$ and $d$ exfoliated on to a rutile or quartz substrate, ellipsometry can be applied for an unambiguous measurement of optical properties and thickness.

\begin{backmatter}
 \bmsection{Funding} This project has received funding from the European Unions's Horizon Europe research and innovation program under grant agreement 101130224 "JOSEPHINE“.

\bmsection{Disclosures} The authors declare no conflicts of interest.

\bmsection{Data availability} Data underlying the results presented in this paper are not publicly available at this time but may be obtained from the authors upon reasonable request.

\bmsection{Supplemental document}
See Supplement 1 for supporting content. 

\end{backmatter}


\bibliography{NKD-Decoupling-Theory_citations}

\begin{thebibliography}{10}
\newcommand{\enquote}[1]{``#1''}

\bibitem{leiGrapheneRecentAdvances2022}
Y.~Lei, T.~Zhang, Y.-C. Lin, \emph{et~al.}, \enquote{Graphene and {{Beyond}}: {{Recent Advances}} in {{Two-Dimensional Materials Synthesis}}, {{Properties}}, and {{Devices}},} {\protect\JournalTitle{ACS Nanoscience Au}} \textbf{2}, 450--485 (2022).

\bibitem{pham2DHeterostructuresUbiquitous2022}
P.~V. Pham, S.~C. Bodepudi, K.~Shehzad, \emph{et~al.}, \enquote{{{2D Heterostructures}} for {{Ubiquitous Electronics}} and {{Optoelectronics}}: {{Principles}}, {{Opportunities}}, and {{Challenges}},} {\protect\JournalTitle{Chemical Reviews}} \textbf{122}, 6514--6613 (2022).

\bibitem{kasparRiseIntelligentMatter2021}
C.~Kaspar, B.~J. Ravoo, W.~G. {van der Wiel}, \emph{et~al.}, \enquote{The rise of intelligent matter,} {\protect\JournalTitle{Nature}} \textbf{594}, 345--355 (2021).

\bibitem{sebastianTwodimensionalMaterialsbasedProbabilistic2022}
A.~Sebastian, R.~Pendurthi, A.~Kozhakhmetov, \emph{et~al.}, \enquote{Two-dimensional materials-based probabilistic synapses and reconfigurable neurons for measuring inference uncertainty using {{Bayesian}} neural networks,} {\protect\JournalTitle{Nature Communications}} \textbf{13}, 6139 (2022).

\bibitem{kwakSensorComputingUsing2023}
D.~Kwak, D.~K. Polyushkin, and T.~Mueller, \enquote{In-sensor computing using a {{MoS2}} photodetector with programmable spectral responsivity,} {\protect\JournalTitle{Nature Communications}} \textbf{14}, 4264 (2023).

\bibitem{mennelUltrafastMachineVision2020}
L.~Mennel, J.~Symonowicz, S.~Wachter, \emph{et~al.}, \enquote{Ultrafast machine vision with {{2D}} material neural network image sensors,} {\protect\JournalTitle{Nature}} \textbf{579}, 62--66 (2020).

\bibitem{migliatomaregaLargescaleIntegratedVector2023}
G.~Migliato~Marega, H.~G. Ji, Z.~Wang, \emph{et~al.}, \enquote{A large-scale integrated vector--matrix multiplication processor based on monolayer molybdenum disulfide memories,} {\protect\JournalTitle{Nature Electronics}} \textbf{6}, 991--998 (2023).

\bibitem{wolffSuperconductingNanowireSinglephoton2020}
M.~A. Wolff, S.~Vogel, L.~Splitthoff, and C.~Schuck, \enquote{Superconducting nanowire single-photon detectors integrated with tantalum pentoxide waveguides,} {\protect\JournalTitle{Scientific Reports}} \textbf{10}, 17170 (2020).

\bibitem{kangExploitingPlasmons2D2022}
L.~Kang, J.~A. Robinson, and D.~H. Werner, \enquote{Exploiting plasmons in {{2D}} metals for refractive index sensing: {{Simulation}} study,} {\protect\JournalTitle{Journal of Applied Physics}} \textbf{132}, 223103 (2022).

\bibitem{pendurthiHeterogeneousIntegrationAtomically2022}
R.~Pendurthi, D.~Jayachandran, A.~Kozhakhmetov, \emph{et~al.}, \enquote{Heterogeneous {{Integration}} of {{Atomically Thin Semiconductors}} for {{Non-von Neumann CMOS}},} {\protect\JournalTitle{Small}} \textbf{18}, 2202590 (2022).

\bibitem{stevesUnexpectedInfraredVisible2020}
M.~A. Steves, Y.~Wang, N.~Briggs, \emph{et~al.}, \enquote{Unexpected {{Near-Infrared}} to {{Visible Nonlinear Optical Properties}} from 2-{{D Polar Metals}},} {\protect\JournalTitle{Nano Letters}} \textbf{20}, 8312--8318 (2020).

\bibitem{lundebergTuningQuantumNonlocal2017}
M.~B. Lundeberg, Y.~Gao, R.~Asgari, \emph{et~al.}, \enquote{Tuning quantum nonlocal effects in graphene plasmonics,} {\protect\JournalTitle{Science}} \textbf{357}, 187--191 (2017).

\bibitem{bankwitzHighqualityFactorTa2O5oninsulator2023a}
J.~R. Bankwitz, M.~A. Wolff, A.~S. Abazi, \emph{et~al.}, \enquote{High-quality factor {{Ta}}{\textsubscript{2}}{{O}}{\textsubscript{5}}-on-insulator resonators with ultimate thermal stability,} {\protect\JournalTitle{Optics Letters}} \textbf{48}, 5783--5786 (2023).

\bibitem{dasTransistorsBasedTwodimensional2021}
S.~Das, A.~Sebastian, E.~Pop, \emph{et~al.}, \enquote{Transistors based on two-dimensional materials for future integrated circuits,} {\protect\JournalTitle{Nature Electronics}} \textbf{4}, 786--799 (2021).

\bibitem{fukamachiLargeareaSynthesisTransfer2023}
S.~Fukamachi, P.~{Sol{\'i}s-Fern{\'a}ndez}, K.~Kawahara, \emph{et~al.}, \enquote{Large-area synthesis and transfer of multilayer hexagonal boron nitride for enhanced graphene device arrays,} {\protect\JournalTitle{Nature Electronics}}  (2023).

\bibitem{radisavljevicSinglelayerMoS2Transistors2011}
B.~Radisavljevic, A.~Radenovic, J.~Brivio, \emph{et~al.}, \enquote{Single-layer {{MoS2}} transistors,} {\protect\JournalTitle{Nature Nanotechnology}} \textbf{6}, 147--150 (2011).

\bibitem{troueExtendedSpatialCoherence2023}
M.~Troue, J.~Figueiredo, L.~Sigl, \emph{et~al.}, \enquote{Extended {{Spatial Coherence}} of {{Interlayer Excitons}} in \$\{{\textbackslash}mathrm\{\vphantom{\}\}}{{MoSe}}\vphantom\{\}\vphantom\{\}\_\{2\}/\{{\textbackslash}mathrm\{\vphantom{\}\}}{{WSe}}\vphantom\{\}\vphantom\{\}\_\{2\}\$ {{Heterobilayers}},} {\protect\JournalTitle{Physical Review Letters}} \textbf{131}, 036902 (2023).

\bibitem{wietekNonlinearNegativeEffective2024}
E.~Wietek, M.~Florian, J.~G{\"o}ser, \emph{et~al.}, \enquote{Nonlinear and {{Negative Effective Diffusivity}} of {{Interlayer Excitons}} in {{Moir}}{\textbackslash}'e-{{Free Heterobilayers}},} {\protect\JournalTitle{Physical Review Letters}} \textbf{132}, 016202 (2024).

\bibitem{dirnbergerMagnetoopticsVanWaals2023}
F.~Dirnberger, J.~Quan, R.~Bushati, \emph{et~al.}, \enquote{Magneto-optics in a van der {{Waals}} magnet tuned by self-hybridized polaritons,} {\protect\JournalTitle{Nature}} \textbf{620}, 533--537 (2023).

\bibitem{grudininHexagonalBoronNitride2023}
D.~V. Grudinin, G.~A. Ermolaev, D.~G. Baranov, \emph{et~al.}, \enquote{Hexagonal boron nitride nanophotonics: A record-breaking material for the ultraviolet and visible spectral ranges,} {\protect\JournalTitle{Materials Horizons}} \textbf{10}, 2427--2435 (2023).

\bibitem{funkeImagingSpectroscopicEllipsometry2016}
S.~Funke, B.~Miller, E.~Parzinger, \emph{et~al.}, \enquote{Imaging spectroscopic ellipsometry of {{MoS}} {\textsubscript{2}},} {\protect\JournalTitle{Journal of Physics: Condensed Matter}} \textbf{28}, 385301 (2016).

\bibitem{wurstbauerImagingEllipsometryGraphene2010}
U.~Wurstbauer, C.~R{\"o}ling, U.~Wurstbauer, \emph{et~al.}, \enquote{Imaging ellipsometry of graphene,} {\protect\JournalTitle{Applied Physics Letters}} \textbf{97}, 231901 (2010).

\bibitem{braeuninger-weimerFastNoncontactWaferScale2018}
P.~{Braeuninger-Weimer}, S.~Funke, R.~Wang, \emph{et~al.}, \enquote{Fast, {{Noncontact}}, {{Wafer-Scale}}, {{Atomic Layer Resolved Imaging}} of {{Two-Dimensional Materials}} by {{Ellipsometric Contrast Micrography}},} {\protect\JournalTitle{ACS Nano}} \textbf{12}, 8555--8563 (2018).

\bibitem{siggerSpectroscopicImagingEllipsometry2022}
F.~Sigger, H.~Lambers, K.~Nisi, \emph{et~al.}, \enquote{Spectroscopic imaging ellipsometry of two-dimensional {{TMDC}} heterostructures,} {\protect\JournalTitle{Applied Physics Letters}} \textbf{121}, 071102 (2022).

\bibitem{rajabpourTunable2DGroupIII2021}
S.~Rajabpour, A.~Vera, W.~He, \emph{et~al.}, \enquote{Tunable {{2D Group-III Metal Alloys}},} {\protect\JournalTitle{Advanced Materials}} \textbf{33}, 2104265 (2021).

\bibitem{nisiLightMatterInteraction2021}
K.~Nisi, S.~Subramanian, W.~He, \emph{et~al.}, \enquote{Light--{{Matter Interaction}} in {{Quantum Confined 2D Polar Metals}},} {\protect\JournalTitle{Advanced Functional Materials}} \textbf{31}, 2005977 (2021).

\bibitem{bankwitzHighqualityFactorTa2O5oninsulator2023}
J.~R. Bankwitz, M.~A. Wolff, A.~S. Abazi, \emph{et~al.}, \enquote{High-quality factor {{Ta}}{\textsubscript{2}}{{O}}{\textsubscript{5}}-on-insulator resonators with ultimate thermal stability,} {\protect\JournalTitle{Optics Letters}} \textbf{48}, 5783--5786 (2023).

\bibitem{fujiwaraSpectroscopicEllipsometryPrinciples2007}
H.~Fujiwara, \emph{Spectroscopic Ellipsometry: Principles and Applications} (John Wiley \& Sons, Chichester, England; Hoboken, NJ, 2007).

\bibitem{britnellElectronTunnelingUltrathin2012}
L.~Britnell, R.~V. Gorbachev, R.~Jalil, \emph{et~al.}, \enquote{Electron {{Tunneling}} through {{Ultrathin Boron Nitride Crystalline Barriers}},} {\protect\JournalTitle{Nano Letters}} \textbf{12}, 1707--1710 (2012).

\bibitem{parkTemperatureDependenceCritical2016}
H.~G. Park, T.~J. Kim, H.~S. Kim, \emph{et~al.}, \enquote{Temperature dependence of the critical points of monolayer {{MoS}}{\textsubscript{2}} by ellipsometry,} {\protect\JournalTitle{Applied Spectroscopy Reviews}} \textbf{51}, 621--635 (2016).

\bibitem{nguyenTemperatureDependenceDielectric2024}
X.~A. Nguyen, L.~V. Le, S.~H. Kim, \emph{et~al.}, \enquote{Temperature dependence of the dielectric function and critical points of monolayer {{WSe2}},} {\protect\JournalTitle{Scientific Reports}} \textbf{14}, 13486 (2024).

\bibitem{luInfluenceElectricField2006}
C.~L. Lu, C.~P. Chang, Y.~C. Huang, \emph{et~al.}, \enquote{Influence of an electric field on the optical properties of few-layer graphene with {{{\emph{AB}}}} stacking,} {\protect\JournalTitle{Physical Review B}} \textbf{73}, 144427 (2006).

\bibitem{martin-sanchezStraintuningOpticalProperties2017}
J.~{Mart{\'i}n-S{\'a}nchez}, R.~Trotta, A.~Mariscal, \emph{et~al.}, \enquote{Strain-tuning of the optical properties of semiconductor nanomaterials by integration onto piezoelectric actuators,} {\protect\JournalTitle{Semiconductor Science and Technology}} \textbf{33}, 013001 (2017).

\bibitem{plumhofStraininducedAnticrossingBright2011}
J.~D. Plumhof, V.~K{\v r}{\'a}pek, F.~Ding, \emph{et~al.}, \enquote{Strain-induced anticrossing of bright exciton levels in single self-assembled {{GaAs}}/{{Al}} x {{Ga}} 1 - x {{As}} and {{In}} x {{Ga}} 1 - x {{As}}/{{GaAs}} quantum dots,} {\protect\JournalTitle{Physical Review B}} \textbf{83}, 121302 (2011).

\bibitem{leeOpticalPropertiesChargedensitywave2008}
K.~E. Lee, C.~I. Lee, H.~J. Oh, \emph{et~al.}, \enquote{Optical properties of the charge-density-wave compound {{CeTe}} 2,} {\protect\JournalTitle{Physical Review B}} \textbf{78}, 134408 (2008).

\bibitem{hilfikerSurveyMethodsCharacterize2008}
J.~N. Hilfiker, N.~Singh, T.~Tiwald, \emph{et~al.}, \enquote{Survey of methods to characterize thin absorbing films with {{Spectroscopic Ellipsometry}},} {\protect\JournalTitle{Thin Solid Films}} \textbf{516}, 7979--7989 (2008).

\bibitem{droletPolynomialInversionSingle1994}
J.-P. Drolet, S.~C. Russev, R.~M. Leblanc, and M.~I. Boyanov, \enquote{Polynomial inversion of the single transparent layer problem in ellipsometry,} {\protect\JournalTitle{Journal of the Optical Society of America A}} \textbf{11}, 3284 (1994).

\bibitem{ibrahimParameterCorrelationComputationalConsiderations1971}
M.~M. Ibrahim and N.~M. Bashara, \enquote{Parameter-{{Correlation}} and {{Computational Considerations}} in {{Multiple-Angle Ellipsometry}}*,} {\protect\JournalTitle{Journal of the Optical Society of America}} \textbf{61}, 1622 (1971).

\bibitem{guAnalyticalMethodDetermine2020}
H.~Gu, S.~Zhu, B.~Song, \emph{et~al.}, \enquote{An analytical method to determine the complex refractive index of an ultra-thin film by ellipsometry,} {\protect\JournalTitle{Applied Surface Science}} \textbf{507}, 145091 (2020).

\bibitem{gilliotExtractionComplexRefractive2012}
M.~Gilliot, \enquote{Extraction of complex refractive index of absorbing films from ellipsometry measurement,} {\protect\JournalTitle{Thin Solid Films}} \textbf{520}, 5568--5574 (2012).

\bibitem{azzamEllipsometryPolarizedLight1987}
R.~M.~A. Azzam and N.~M. Bashara, \emph{Ellipsometry and Polarized Light} ({North-Holland : Sole distributors for the USA and Canada, Elsevier Science Pub. Co.}, Amsterdam; New York, 1987).

\bibitem{haugeDesignOperationETA1973}
P.~S. Hauge and F.~H. Dill, \enquote{Design and {{Operation}} of {{ETA}}, an {{Automated Ellipsometer}},} {\protect\JournalTitle{IBM Journal of Research and Development}} \textbf{17}, 472--489 (1973).

\bibitem{jellisonMeasurementOpticalFunctions1997}
G.~E. Jellison, F.~A. Modine, and L.~A. Boatner, \enquote{Measurement of the optical functions of uniaxial materials by two-modulator generalized ellipsometry: Rutile ({{TiO}}\_2),} {\protect\JournalTitle{Optics Letters}} \textbf{22}, 1808 (1997).

\bibitem{berremanOpticsStratifiedAnisotropic1972}
D.~W. Berreman, \enquote{Optics in {{Stratified}} and {{Anisotropic Media}}: 4{\texttimes}4-{{Matrix Formulation}},} {\protect\JournalTitle{Journal of the Optical Society of America}} \textbf{62}, 502 (1972).

\bibitem{schubertInfraredEllipsometrySemiconductor2004}
M.~Schubert, \emph{Infrared Ellipsometry on Semiconductor Layer Structures: Phonons, Plasmons, and Polaritons}, no. v. 209 in Springer Tracts in Modern Physics (Springer, Berlin ; New York, 2004).

\bibitem{newvilleLMFITNonLinearLeastSquare2014}
M.~Newville, T.~Stensitzki, D.~B. Allen, and A.~Ingargiola, \enquote{{{LMFIT}}: {{Non-Linear Least-Square Minimization}} and {{Curve-Fitting}} for {{Python}},} Zenodo (2014).

\bibitem{hilfikerDeterminingThicknessRefractive2017}
J.~N. Hilfiker, M.~Stadermann, J.~Sun, \emph{et~al.}, \enquote{Determining thickness and refractive index from free-standing ultra-thin polymer films with spectroscopic ellipsometry,} {\protect\JournalTitle{Applied Surface Science}} \textbf{421}, 508--512 (2017).

\bibitem{devoreRefractiveIndicesRutile1951}
J.~R. DeVore, \enquote{Refractive {{Indices}} of {{Rutile}} and {{Sphalerite}},} {\protect\JournalTitle{Journal of the Optical Society of America}} \textbf{41}, 416 (1951).

\bibitem{okanoVoltageControlledDielectricFunction2020}
S.~Okano, A.~Sharma, F.~Ortmann, \emph{et~al.}, \enquote{Voltage-{{Controlled Dielectric Function}} of {{Bilayer Graphene}},} {\protect\JournalTitle{Advanced Optical Materials}} \textbf{8}, 2000861 (2020).

\bibitem{polyanskiyRefractiveindexinfoDatabaseOptical2024}
M.~N. Polyanskiy, \enquote{Refractiveindex.info database of optical constants,} {\protect\JournalTitle{Scientific Data}} \textbf{11}, 94 (2024).

\bibitem{aspnesDielectricFunctionsOptical1983}
D.~E. Aspnes and A.~A. Studna, \enquote{Dielectric functions and optical parameters of {{Si}}, {{Ge}}, {{GaP}}, {{GaAs}}, {{GaSb}}, {{InP}}, {{InAs}}, and {{InSb}} from 1.5 to 6.0 {{eV}},} {\protect\JournalTitle{Physical Review B}} \textbf{27}, 985--1009 (1983).

\bibitem{weberOpticalConstantsGraphene2010}
J.~W. Weber, V.~E. Calado, and M.~C.~M. Van De~Sanden, \enquote{Optical constants of graphene measured by spectroscopic ellipsometry,} {\protect\JournalTitle{Applied Physics Letters}} \textbf{97}, 091904 (2010).

\bibitem{ermolaevBroadbandOpticalProperties2020}
G.~A. Ermolaev, Y.~V. Stebunov, A.~A. Vyshnevyy, \emph{et~al.}, \enquote{Broadband optical properties of monolayer and bulk {{MoS2}},} {\protect\JournalTitle{npj 2D Materials and Applications}} \textbf{4}, 21 (2020).

\bibitem{ghoshDispersionequationCoefficientsRefractive1999}
G.~Ghosh, \enquote{Dispersion-equation coefficients for the refractive index and birefringence of calcite and quartz crystals,} {\protect\JournalTitle{Optics Communications}} \textbf{163}, 95--102 (1999).

\bibitem{liMicrowaveBirefringentMetamaterials2016}
Y.~Li, J.~Zhang, H.~Ma, \emph{et~al.}, \enquote{Microwave birefringent metamaterials for polarization conversion based on spoof surface plasmon polariton modes,} {\protect\JournalTitle{Scientific Reports}} \textbf{6}, 34518 (2016).

\bibitem{schmidtAnisotropicBruggemanEffective2013}
D.~Schmidt and M.~Schubert, \enquote{Anisotropic {{Bruggeman}} effective medium approaches for slanted columnar thin films,} {\protect\JournalTitle{Journal of Applied Physics}} \textbf{114}, 083510 (2013).

\end{thebibliography}






\end{document}